\documentstyle[twoside,fleqn,npb,epsfig]{article}
\newcommand{\AmS}{{\protect\the\textfont2
  A\kern-.1667em\lower.5ex\hbox{M}\kern-.125emS}}

\def\ds{D^{\ast}}
\def\d0{D^{0}}

\def\dsk3pi{ {\ds}^{+} \rightarrow \d0 \pi^{+}_{S}%
        \rightarrow (K^{-} \pi^{+} \pi^{+} \pi^{-}) \pi^{+}_{S} }

\def\g2{GeV$^{2}$}

\title{Charm Photoproduction in ep Collisions at HERA }
 
\author{Yehuda Eisenberg%                                           
 \address{Weizmann Institute of Science, Particle Physics
  Department,  
        Rehovot, ISRAEL \\ email: yehuda@mail.desy.de}%
        \thanks{Supported by the German Israeli Foundation and 
                by the Minerva Foundation.}\\
             On behalf of the ZEUS Collaboration}
 
\begin{document}
                                
\begin{abstract}
%%A short abstract, please.
 
We report the latest results of the ZEUS collaboration on the 
photoproduction of $D^*$ mesons in a low $W$ range.     
Differential cross sections as function of  $p_{\perp}^{\ds}$ and
$\eta^{\ds}$ are measured and 
 compared with several recent NLO pQCD calculations. 
The differential cross-sections in a restricted kinematical region are higher 
than the NLO calculations, in particular in the forward
 (proton) direction. A recent pQCD model (BKL) describes the data reasonably well.
 
\end{abstract}
%%%%%%%%%%%%%%%%%%%%%%%%%%
 
\maketitle
 
\section{INTRODUCTION}

Heavy quark photoproduction
can be used to probe
            pQCD calculations
with a hard scale given by     
the heavy quark mass and the high transverse momentum          
of the produced parton                                   
   ($m_Q\gg \Lambda_{QCD}$).             
Two types of NLO calculations with different approaches are available
for comparison with measurements of charm photoproduction at HERA.
The massive charm approach~\cite{MassC} assumes light quarks to be 
the only active flavours within the structure functions of
the proton and the photon, while the massless charm
approach~\cite{LessC.K,LessC.C} 
also treats charm as an active flavour and is thus only  
 valid for $p_{\perp}\gg m_c$.
 
The data taken by the         
ZEUS collaboration during 1996/1997 corresponds to an integrated luminosity
of about $37\,\mbox{pb}^{-1}$. In a subsample of about $17\,\mbox{pb}^{-1}$
%a small calorimeter positioned 44m from the interaction point in the electron direction
a small calorimeter positioned along the beam pipe                                
was used to tag low $W$ events,
 $80$\, $ < W_{\gamma p} < 120$\,GeV. The results of the high $W$ region
($130$\, $ < W_{\gamma p} < 280$\,GeV) have been published before~\cite{dstar98} and
will not be shown here. This is the first presentation of our low $W$ results.
Charm was identified by the observation of                        
 $D^*$(2010) mesons, which were
reconstructed in the following decay modes:
$D^{*+}\rightarrow D^0 \pi_s^+ \rightarrow ( K^- \pi^+ ) \pi_s^+$ $(Br=0.0262 \pm 0.0010)$
and
$D^{*+}\rightarrow D^0 \pi_s^+ \rightarrow ( K^- \pi^+ \pi^+ \pi^-)\pi_s^+$ $(Br=0.051\pm0.003)$     
and charge conjugates.                                    
The  kinematic range studied was
 $p_{\perp}^{\ds} > 2$ GeV and 
  $-1.5 < \eta^{\ds} < 1.5$ for the high $W$ region, and $2 < p_{\perp}^{\ds} < 8\,$ GeV
%$p_{\perp}^{\ds} =\mbox{2--8}\,$GeV and $-1.0 < \eta^{\ds} < 1.5$ for the low $W$ region.
 and $-1.0 < \eta^{\ds} < 1.5$ for the low $W$ region.
The pseudorapidity is $\eta^{\ds} = -\ln(\tan\frac{\theta}{2})$, where
$\theta$ is the polar angle with respect to the proton beam direction.
 
Charged tracks were measured in the central tracking detector.         
Cross sections were calculated in the photoproduction range of photon
virtualities $Q^2 < 1 \,\mbox{GeV}^2$ ( $Q^2 < 0.015\,\mbox{GeV}^2$ for the tagged data).

%%%%%%%%%%%%%%%%%%%%%%%%%%%%%%%%%%%%%%%

\section{ $\ds$ RECONSTRUCTION}
 
%%%%%%%%%%%%%%%%%%%%%%%%%%%%%%%%%%%%%%%%%%%%                      
 
$\ds$ events have been selected by                      
means of the mass difference ($\Delta M$) method~\cite{dmMethod}.       
In the high $W$ region we have observed~\cite{dstar98}
 $3702\pm 136$ $D^*$'s in the $D^0 \rightarrow ( K \pi )$
 decay mode with $p_{\perp}^{\ds} > 2\,$GeV, and $1397\pm 108$ in the $(K\pi\pi\pi)$
decay mode with $p_{\perp}^{\ds} > 4\,$GeV ($M(D^0) = 1.80$--$1.92\,$GeV).               
 In the low $W$ region we triggered only
the $(K\pi)$ decay mode, and observed $550\pm 36$ $\ds$ events in the range
%$p_{\perp}^{\ds} = 2$--$8\,$GeV (Fig.~1).                           
   $2 < p_{\perp}^{\ds} < 8\,$ GeV (Fig.~1).                           
%Since we did not have particle identification, all tracks were assumed
  All tracks were assumed
to be pions and kaons in turn; wrong charge $\ds$ combinations~\cite{dstar98}            
 were used as a background
distribution (dashed curve in Fig.~1), normalized outside the signal region.
%% Fig. 1
     \begin{figure}    
 \psfig{figure=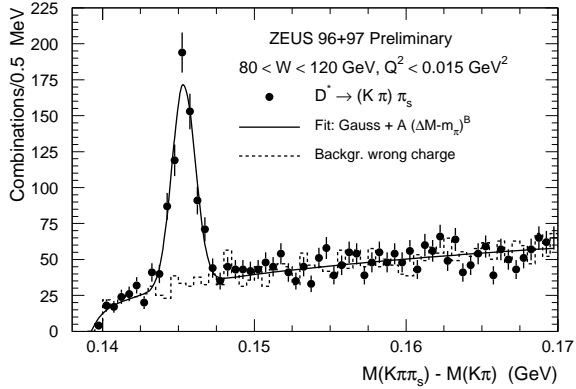,width=\hsize}% 
     \caption{$\Delta M$ distribution for the events inside the $D^o$ mass
      region. The wrong charge distribution is shown as a dashed histogram.}
     \end{figure}

%%%%%%%%%%%%%%%%%%%%%%%%%%%%%%%%%%%%%%%
 
\section{$\ds$ CROSS SECTIONS AND COMPARISON WITH CALCULATIONS} 
 
The $\ds$ differential cross sections in the low $W$ region are shown in Fig.~2 to~5.
For comparing the experimental data to the NLO QCD calculations,
we have used the $\ds$ branching value measured by OPAL~\cite{opal}
   {$f(c \rightarrow D^{*+} + ...) = 0.222\pm 0.014\pm 0.014$}.
For the charm fragmentation to $\ds$ the Peterson~\cite{peterson} fragmentation
function was used:                           
$${D_c(z) = N \frac{z(1-z)^2}{[(1-z)^2 + {\epsilon}z]^2}\;,\qquad          
     z=\frac{p_{D^*}}{p_c}}\;. $$    
In the massive calculation $\epsilon = 0.036$ was obtained from a recent fit of Nason and Oleari%
~\cite{nason99} to ARGUS data.
Alternatively, the Peterson fragmentation was replaced by fragmentation 
effects estimated by a leading order Monte Carlo (Pythia).                          
Initial and final state radiation were not included.
The results of both calculations for the low $W$ region 
 are shown in Figs.~2 and~3.
%% figure 2
     \begin{figure}    
 \psfig{figure=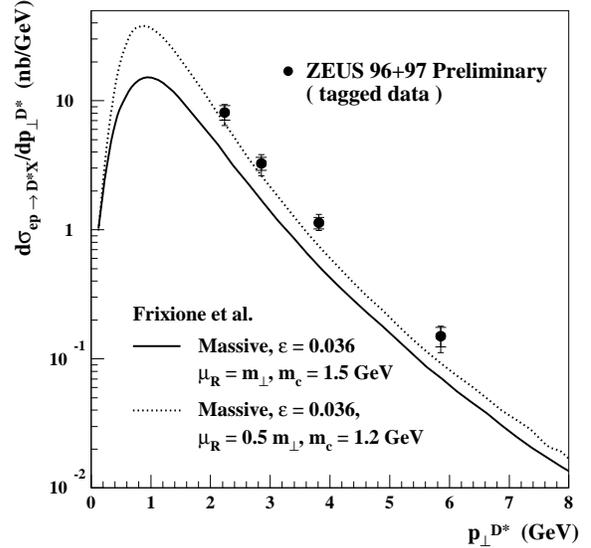,width=\hsize}\par
     \caption{ZEUS tagged data compared to massive NLO calculations of   
      ${d\sigma}/{dp_{\perp}^{\ds}}$.}  
     \end{figure}    
The cross sections are compared with                           
 NLO QCD massive calculations using MRSG for the proton structure function (SF)
%the photon structure function GRV-G HO. Again the pythia fragmentation is
%the dashed lines (Fig.~3). A clear indication that the theoretical massive calculation
and GRV-G HO for the photon.                     
                         The theoretical massive calculations
are below the data, in particular in the forward (proton) direction, although
the Pythia fragmentation slightly improves the agreement.
 
A comparison with massless calculations~\cite{LessC.K},
 which are expected to become valid mainly at
higher $p_{\perp}^{\ds}$, is shown in Fig.~4 for several photon structure functions.
Some sensitivity to the photon SF seems to be present, but the excess in the forward
direction is evident. The structure function GS-G-96 HO describes the data best.
 
Recently Berezhnoy, Kiselev and Likhoded (BKL) have suggested a new model for
describing $\ds$ photoproduction~\cite{BKL}. In this  
 tree level pQCD $O(\alpha \alpha_s^3)$ calculation, they       
hadronize the ($c,\bar q$) state produced in pQCD, taking into   
%%account higher twist terms of $\mathcal{O}(\frac{1}{p_\perp})$ at
account higher twist terms at
$p_\perp \simeq m_c$.                                                  
  Thus the model is supposed to be valid over the whole $p_\perp$ range studied.
  No explicit resolved component is used.  
   Singlet and octet color states both contribute to $D^*$ production.  
The color state ratio $O(8)/O(1)$ is a free parameter in this model and was  
  tuned to the ZEUS untagged results~\cite{dstar98}, yielding a value of 1.3. 
 
  Comparison of these calculations, for the same Octet/Singlet mixture, with the ZEUS
  tagged low $W$ data is shown in Fig.~5. A better agreement with the data is observed 
 than that for the NLO calculations.
%% figure 3
\begin{figure}
\psfig{figure=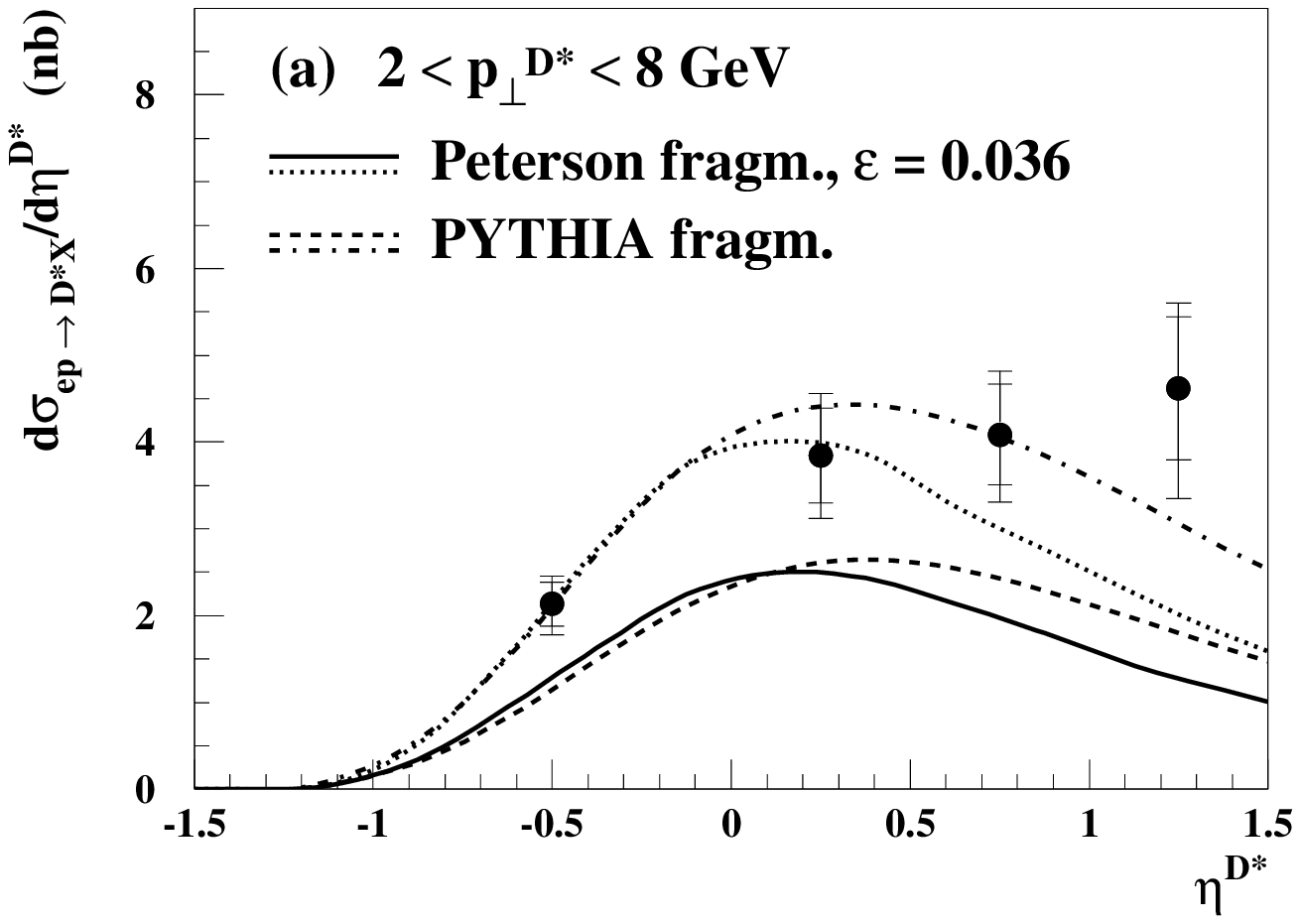,width=\hsize}      
  \psfig{figure=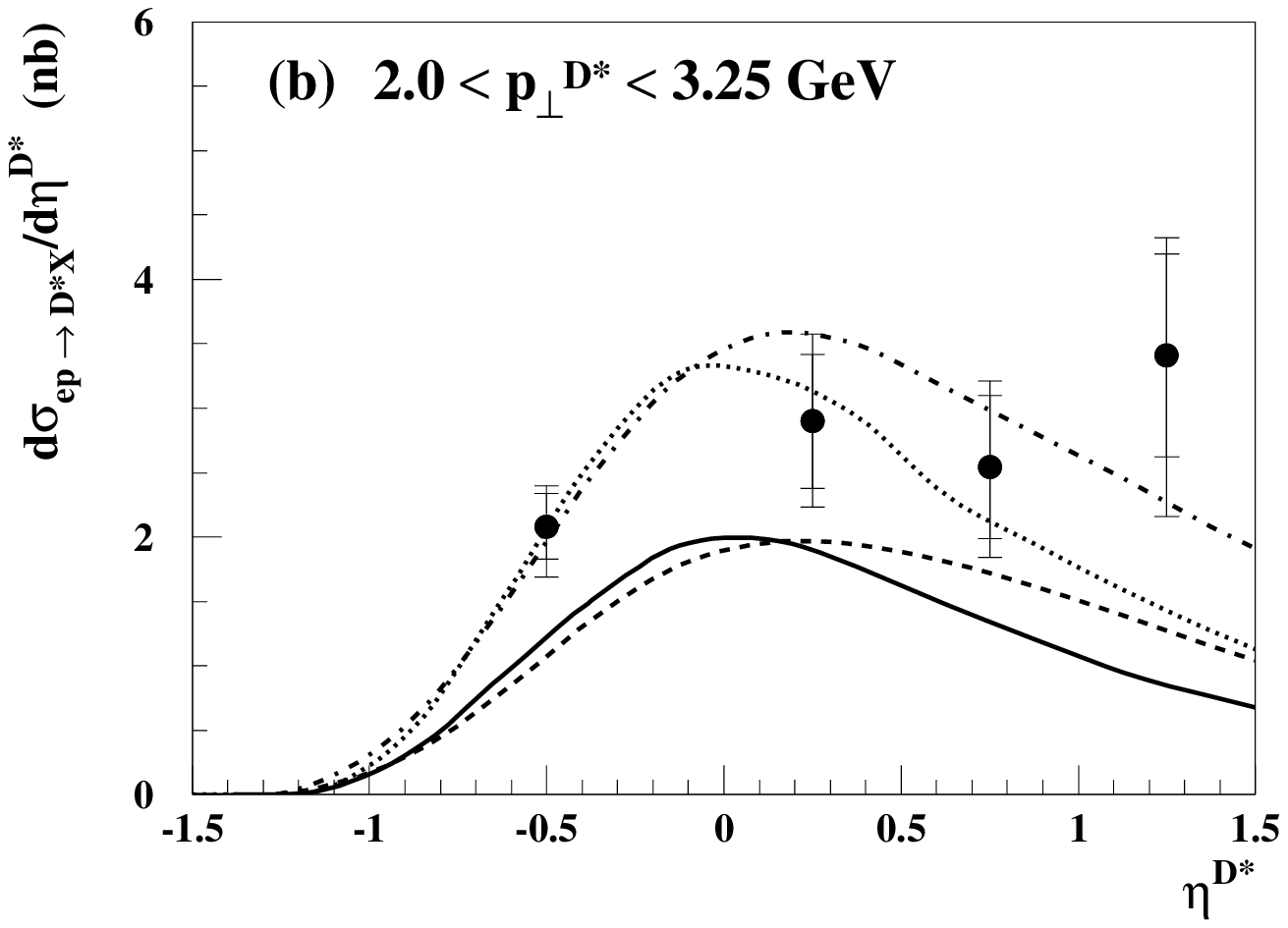,width=\hsize}  
  \psfig{figure=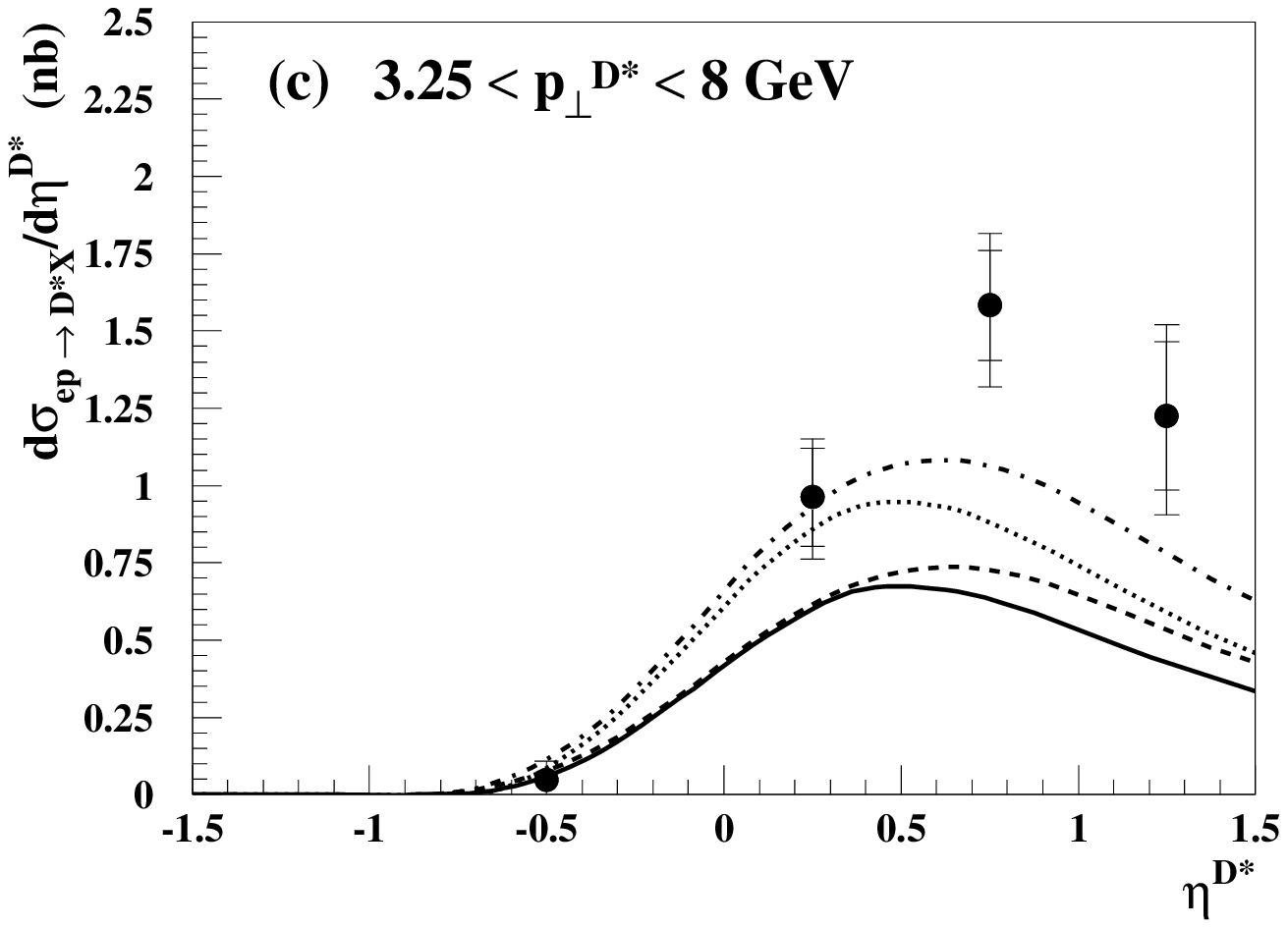,width=\hsize}    
     \caption{ZEUS differential cross sections ${d\sigma}/{d\eta}$     
        compared to the massive NLO predictions. The dashed/dashed-dotted lines
            correspond to the Pythia fragmentation. 
       Parameter sets are as in Fig.2.}
\end{figure}
%% figure 4
\begin{figure}
  \psfig{figure=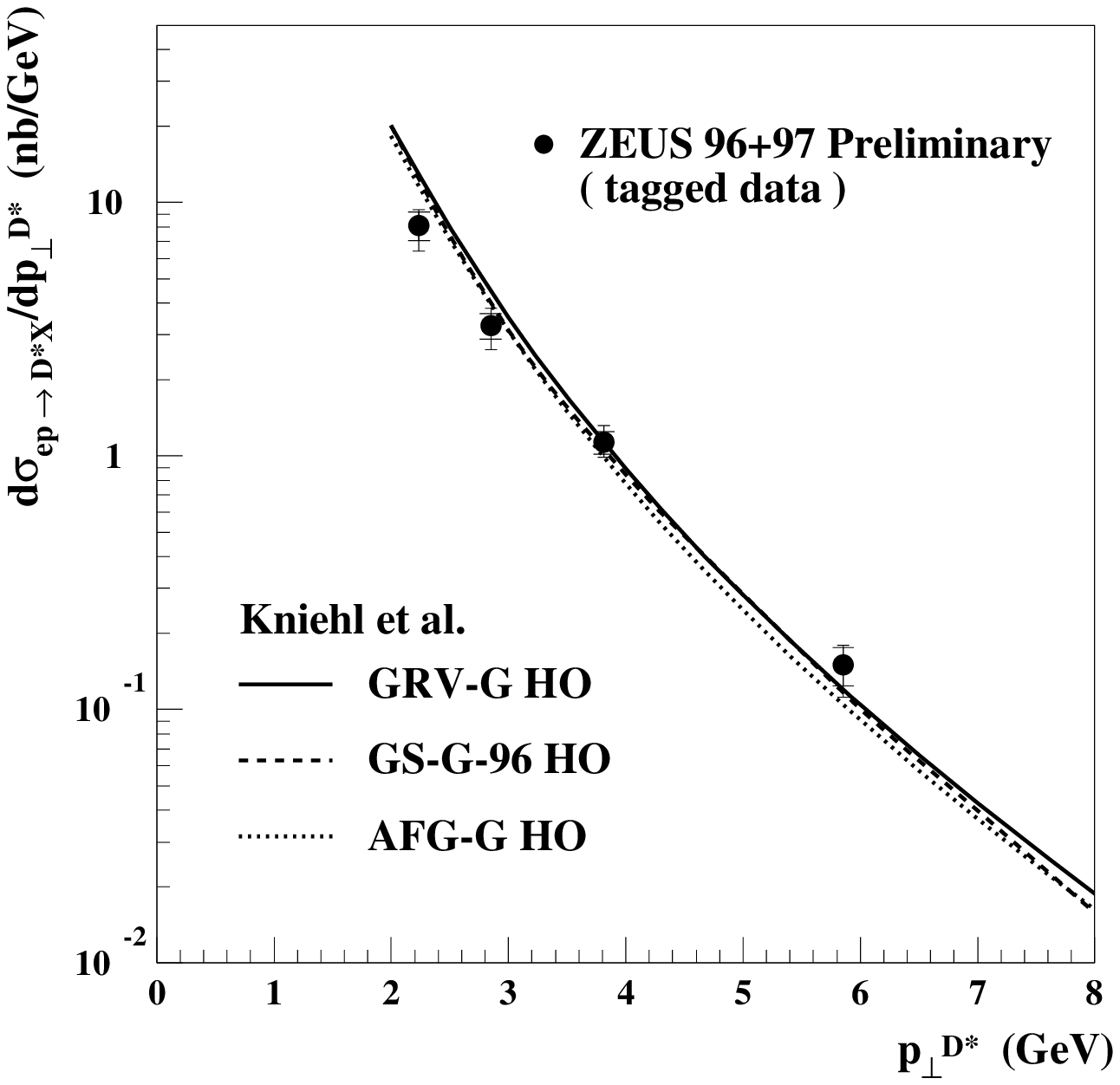,width=\hsize}                    
  \psfig{figure=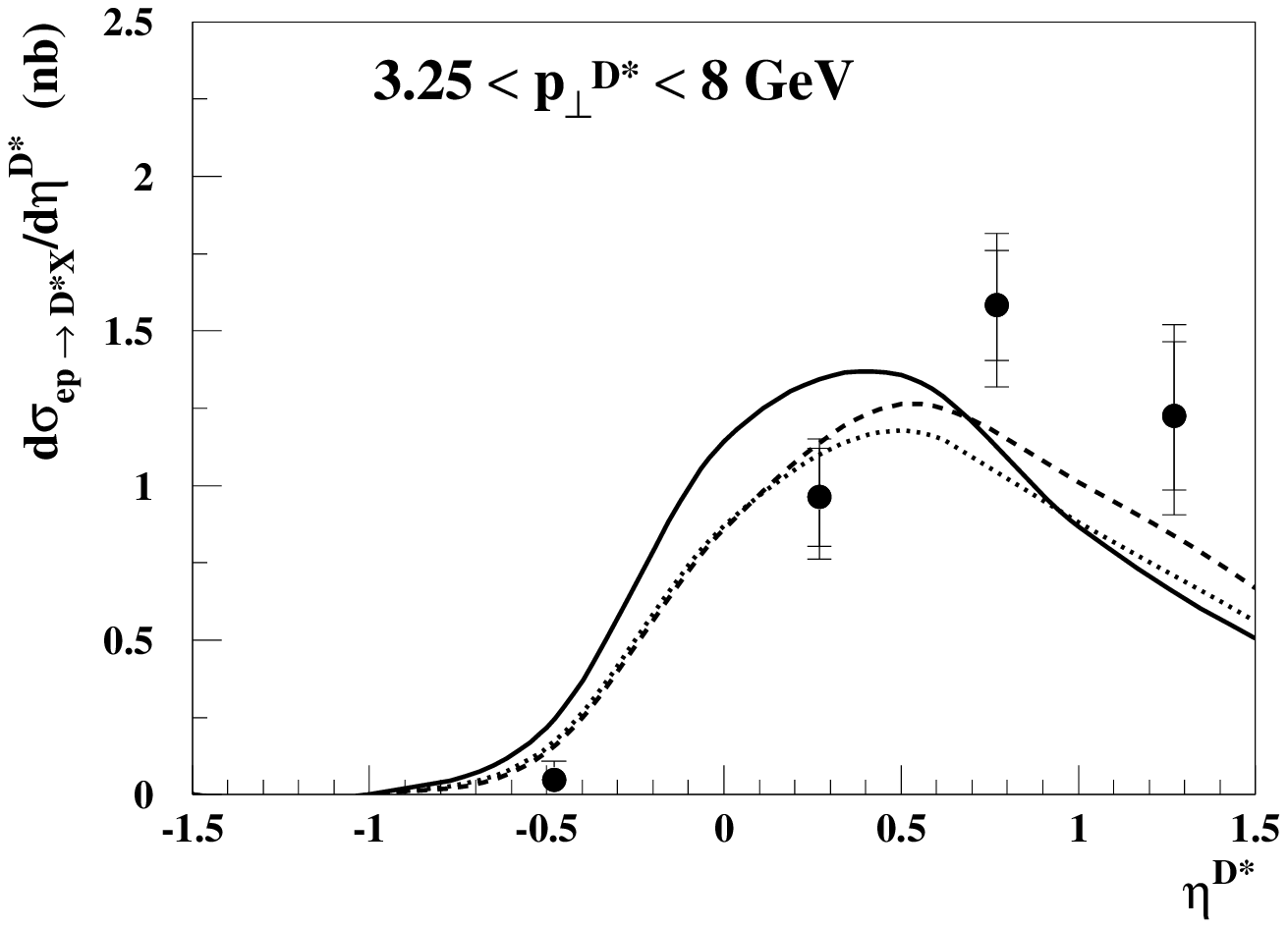,width=\hsize}   
     \caption{ZEUS data and massless NLO predictions~\cite{LessC.K} of differential cross   
       sections for several photon structure functions. Peterson fragmentation was
     used with $\epsilon = 0.116$.} 
\end{figure}
%% Fig. 5    
\begin{figure}
  \psfig{figure=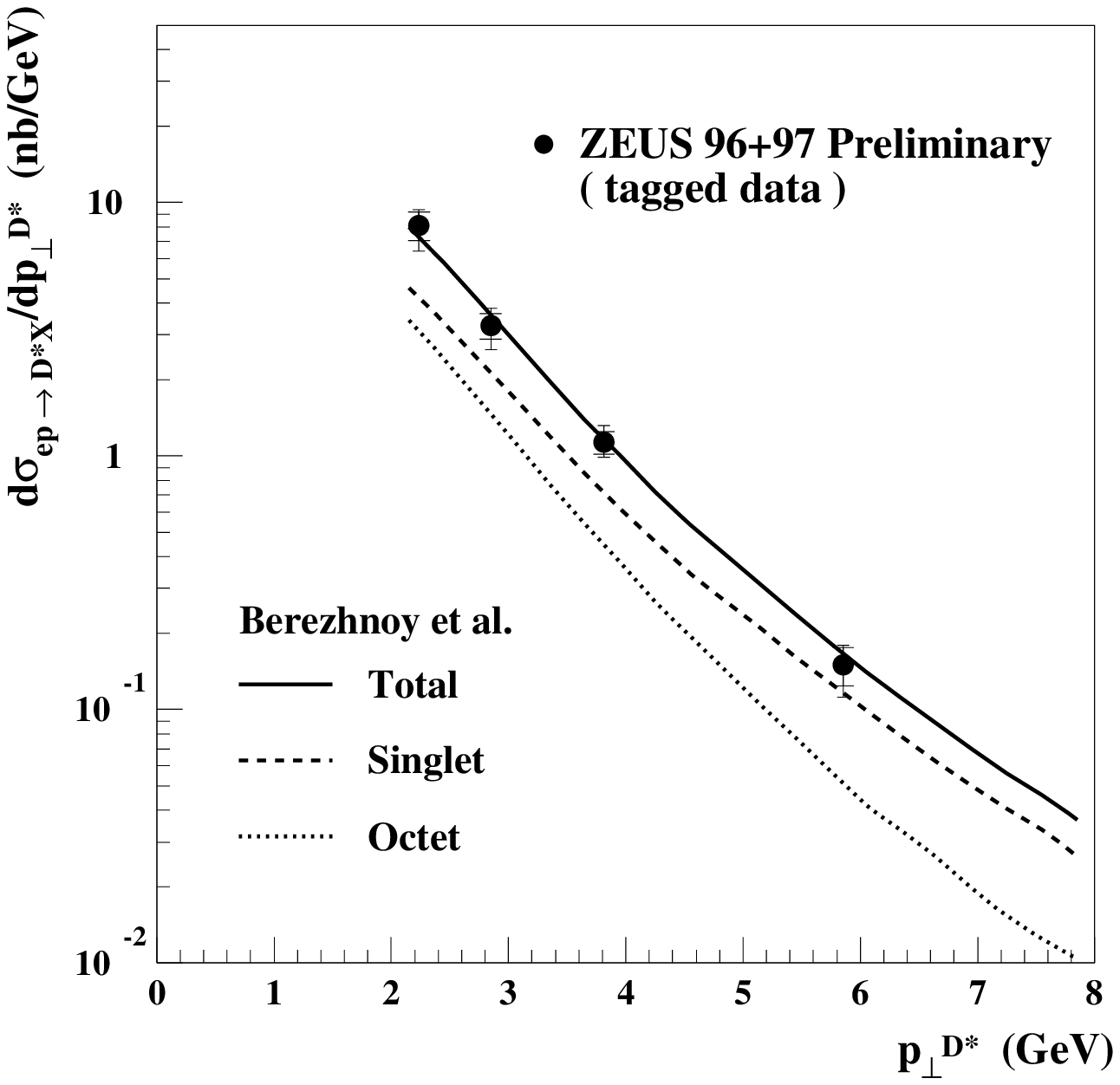,width=\hsize}\par                 
  \psfig{figure=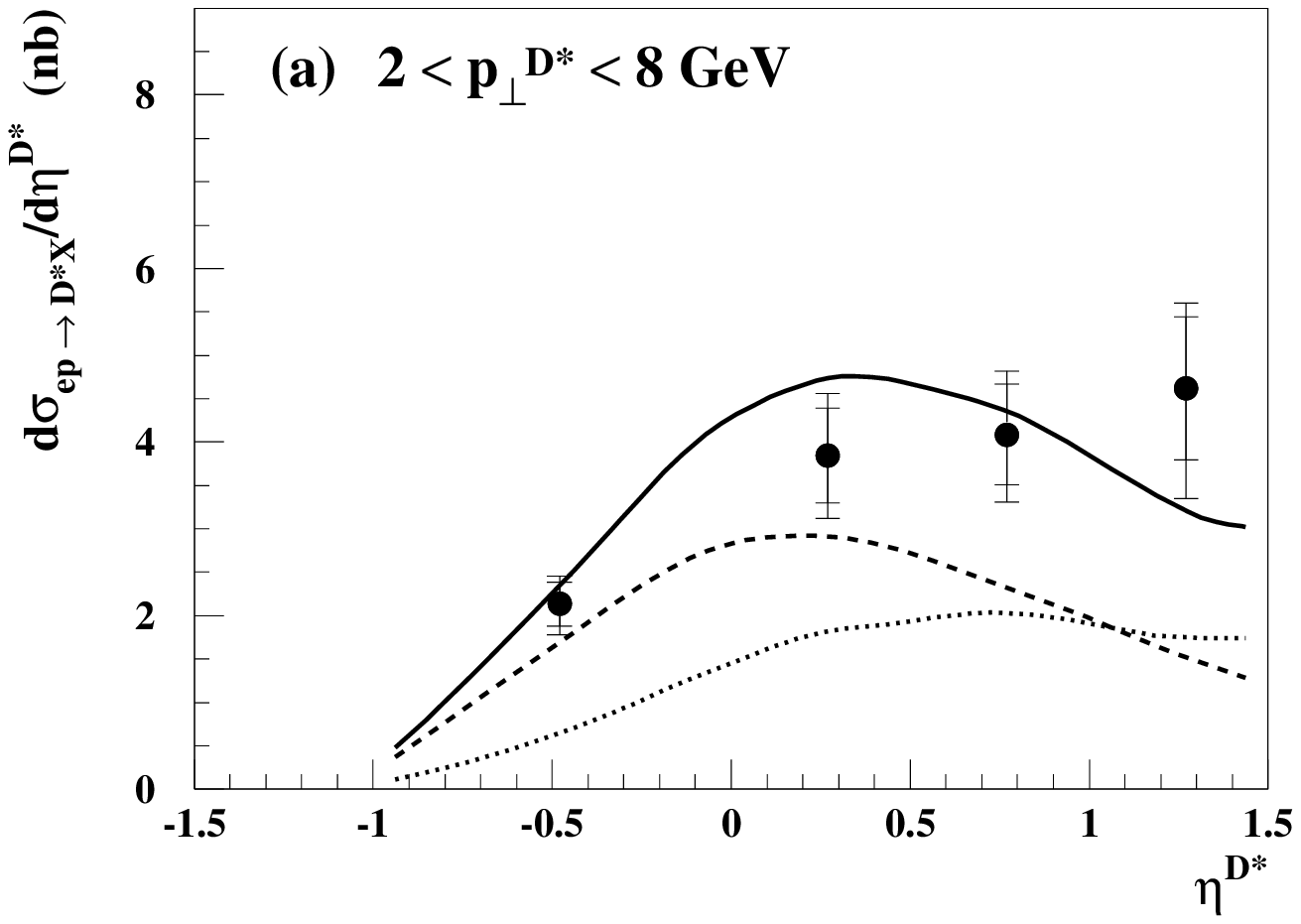,width=\hsize}   
     \caption{Comparison of the ZEUS low $W$ data with the BKL model. 
          The Octet/Singlet ratio is 1.3, as tuned for the high $W$ region.}
\end{figure}
 
\newpage
%
% ---- Bibliography ----
%

\end{document}